\documentclass[journal=jacsat,manuscript=article]{achemso}
\usepackage{soul}
\usepackage{xcolor}
\usepackage{amssymb}
\sethlcolor{red}

\usepackage[version=3]{mhchem}

\author{Shashwat Anand}
\affiliation{Materials Sciences Division, Lawrence Berkeley National Laboratory, CA, USA}

\author{Tara P. Mishra}
\affiliation{Materials Sciences Division, Lawrence Berkeley National Laboratory, CA, USA}

\author{Peichen Zhong}
\affiliation{Materials Sciences Division, Lawrence Berkeley National Laboratory, CA, USA}
\alsoaffiliation[Second University]
{Department of Material Science and Engineering, University of California, Berkeley, CA, USA}

\author{Yunyeong Choi}
\affiliation{Department of Material Science and Engineering, University of California, Berkeley, CA, USA}

\author{KyuJung Jun}
\affiliation{Materials Sciences Division, Lawrence Berkeley National Laboratory, CA, USA}
\alsoaffiliation[Second University]
{Department of Material Science and Engineering, University of California, Berkeley, CA, USA}

\author{Tucker Holstun}
\affiliation{Materials Sciences Division, Lawrence Berkeley National Laboratory, CA, USA}
\alsoaffiliation[Second University]
{Department of Material Science and Engineering, University of California, Berkeley, CA, USA}

\author{Gerbrand Ceder}
\email{gceder@berkeley.edu}
\affiliation{Materials Sciences Division, Lawrence Berkeley National Laboratory, CA, USA}
\alsoaffiliation[Second University]
{Department of Material Science and Engineering, University of California, Berkeley, CA, USA}

\title[An \textsf{achemso} demo]
  {Origin of Enhanced Performance when Mn-Rich Rocksalt Cathodes transform to $\delta$-DRX}
\abbreviations{IR,NMR,UV}

\begin{document}

\begin{abstract}
Most Mn-rich cathodes are known to undergo phase transformation into structures resembling spinel-like ordering upon electrochemical cycling. Recently, the irreversible transformation of Ti-containing Mn-rich disordered rock-salt cathodes into a phase --- named $\delta$ --- with nanoscale spinel-like domains has been shown to increase energy density, capacity retention, and rate capability.  However, the nature of the boundaries between domains and their relationship with composition and electrochemistry are not well understood. In this work, we discuss how the transformation into the multi-domain structure results in eight variants of Spinel domains, which is crucial for explaining the nanoscale domain formation in the $\delta$-phase. We study the energetics of crystallographically unique boundaries and the possibility of Li-percolation across them with a fine-tuned CHGNet machine learning interatomic potential. Energetics of $16d$ vacancies reveal a strong affinity to segregate to the boundaries, thereby opening Li-pathways at the boundary to enhance long-range Li-percolation in the $\delta$ structure. Defect calculations of the relatively low-mobility Ti show how it can influence the extent of Spinel ordering, domain morphology and size significantly; leading to guidelines for engineering electrochemical performance through changes in composition.

\end{abstract}

\section{Introduction}

Mn is of particular interest as a redox-active element in cathode materials due to its low cost, low toxicity and high natural abundance.\cite{Lee2018Nature_Mn2_Mn4, Lun2020_Chem_Mn_DRX, olivetti2017lithium} In a cathode at a high state of change,  Mn$^{4+}$ ions form which is a stable valence state less prone to oxygen loss than Ni$^{4+}$ or Co$^{4+}$ such redox stability is expected to enhance the thermal stability of the material which can be exploited to achieve energy gains at the pack level\cite{lu2001layered}. This makes Mn-based chemistries among the most exciting and promising directions to lower the cost of Li-ion cells while maintaining or enhancing energy density. However, commercialized NMC chemistries rely on only \ce{Ni^{2+/4+}} and \ce{Co^{3+/4+}} redox, leaving \ce{Mn^{4+}}as an inactive stabilizer, increasing cost and limiting thermal stability. Mn-redox has been studied for cathode applications in many structures, including spinel \cite{thackeray1984electrochemical}, layered \cite{PeterBruce1996_Nature_RechargableCathode}, orthorhombic\cite{Thackeray1993_MatResBull_Ortho_to_Spinel}, olivine \cite{li2002limnpo4}, and the rock-salt structure\cite{Pralong2016_NatMater_Li4Mn2O5, Lee2018Nature_Mn2_Mn4, Lun2020_Chem_Mn_DRX} However, due to the mobile nature of low valent Mn, all close packed oxides with high Mn-content transform irreversibly into structures resembling spinel-like ordering upon electrochemical cycling.\cite{Reed2001_Layered_to_Spinel_ECSSLett, Kristin2014_AEM_Layered_to_Spinel, YMChiang_2000_Layered_to_Delta_ChemMater, Thackeray1993_MatResBull_Ortho_to_Spinel, jang1999_JECS_Ortho_to_Spinel,Zijian_2024_Nature_Energy, YMChiang1999_ECSSLett_Origin_of_Transformation} Even in layered Li and Mn-rich cathodes, where all Mn is oxidized to 4+ at synthesis, there is evidence of slow changes to spinel-like arrangements after oxygen loss and associated Mn reduction.\cite{Nayak2014_JECS_Layered_Li-excess, Shimoda2017_JMCA_Layered_to_Spinel, Kristin2014_AEM_Layered_to_Spinel} In addition to the mobile nature of Mn, the phase transformation can be attributed to the strong driving force to form the spinel structure which is well-known to be the ground state at the \ce{LiMn2O4} composition.\cite{Mishra1999_PRB_Structural_Stability,Paulsen1999_ChemMaterLiMnO_PD} 

Irreversible transition metal migration and phase transformations are generally considered detrimental for cathode materials as they may trigger voltage hysteresis,\cite{Lyu2015_ChemMater_Hysterisis} poor kinetics\cite{Croy2015_AccountsChemRes_Review} and capacity degradation\cite{Komaba_2010_EChemCom_Degradation}. However, it was recently shown that the transformation of the Mn-rich disordered rocksalt (DRX) (\ce{Li_{1.05}Mn_{0.85}Ti_{0.1}O_2}) into structure with spinel-like ordering --- named the $\delta$-phase --- \cite{Zijian_2024_Nature_Energy} exhibits high energy density, improved rate capability and negligible voltage fade upon cycling.\cite{HanMingHau2024earth} The ordering in the $\delta$-phase has very recently been shown to have a short nanoscale coherence length, meaning that the definition of the spinel ordered cation sub-lattice (the $16d$ Wyckoff site) changes every few nanometers giving rise to a multi-domain structure where each domain can be characterized by a ``variant" of Spinel (or a $16c/16d$ site occupancy on the cubic lattice) that is distinct from its neighbors.\cite{HanMingHau2024earth} The unique nanostructure of the $\delta$-phase is helpful in achieving much better rate capability than DRX \cite{Zijian_2024_Nature_Energy,HanMingHau2024earth,Tucker_2024_Pulsing_AM} and suppressing the two-phase cubic-to-tetragonal phase transformation which limits the usable capacity of well-ordered Spinel.\cite{van2000phase,erichsen2020tracking,thackeray1984electrochemical} 

\begin{figure}[hbt!]
\centering
\includegraphics[scale=0.2]{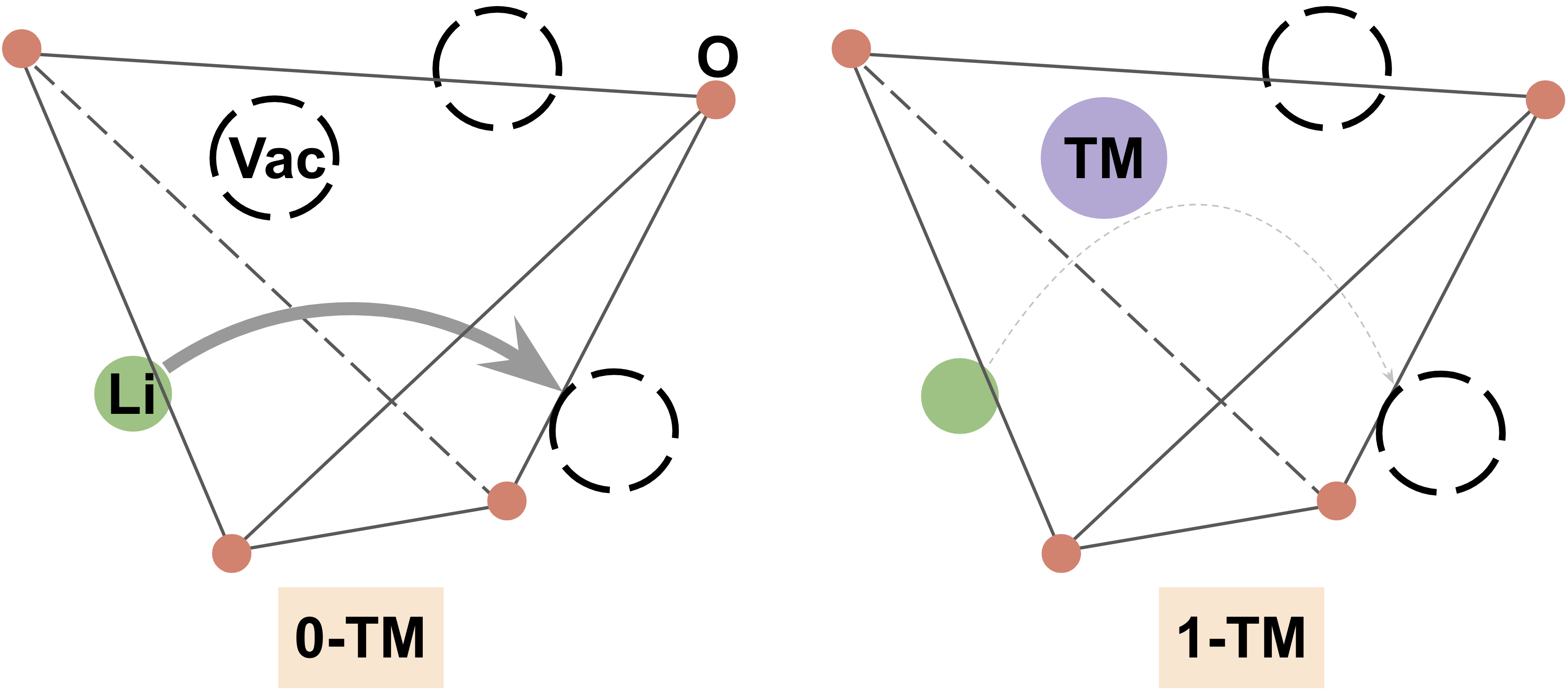}
\caption{Two examples of possible tetrahedral Li-diffusion pathways (0-TM and 1-TM) for Li-diffusion in Spinel and disordered rocksalt (DRX) structures. The tetrahedral pathway are formed by Oxygen atoms (small orange) in its corners and face share with four nearest neighbor octahedral sites around it which can be occupied by a Li (small green sphere), a transition metal (TM, large purple circle) atom or remain vacant (Vac, dashed circle).}
\label{fig:0_TM_1_TM_Examples}
\end{figure}

Li transport most close-packed oxides occurs through an oct-tet-oct mechanism. Figure \ref{fig:0_TM_1_TM_Examples} shows two examples of the tetrahedral pathway, which face-shares with 4 octahedral sites around it. The face-sharing octahedral sites can be occupied by a Li, a transition metal (TM) atom or remain vacant. The tetrahedral pathways are named as ``$n$-TM'' (0-TM and 1-TM) where $n$ refers to the number of octahedral transition metal (TM) atoms that are face-sharing with the tetrahedral site. In general, the migration barriers for Li-diffusion are inversely correlated to $n$ and Li-transport relies significantly on the long-range connectivity of the low-barrier 0-TM channels.\cite{Urban2014_AEM, Lee2014_Science_DRX} In multi-domain spinel-like nanostructures such as the $\delta$-phase however, the nature of cation ordering at the boundaries between domains of different Spinel variants and their impact on Li-transport are not well-understood yet. To the best of our knowledge, the number of possible Spinel variants which emerge upon transforming to multi-domain Spinel has not been discussed previously in the context of Mn-rich cathode materials. It is well-understood that the parent rocksalt lattice can be divided into two sublattices ($16c$ and $16d$) of the Spinel structure which straightforwardly accounts for two Spinel variants; where Mn occupies $16c$ in one and $16d$ in the other. Microscopy images of the transformed nanostructures show regions however, where the spinel-like domains form tri-junctions between each other,\cite{YMChiang_2000_Layered_to_Delta_ChemMater} which can only be explained if more than two variants exist. A more detailed crystallographic understanding of the DRX-to-$\delta$ transformation is therefore necessary before studying the influence of boundaries on Li transport. Furthermore, the impact of Ti and Li-content on the domain morphology of the $\delta$-phase and the resulting suppression of the two-phase cubic-to-tetragonal transformation is not well-understood. In this work, we construct the crystallographically unique low-index domain boundaries possible in the $\delta$-phase and study their effect on its electrochemical properties using a fine-tuned CHGNet\cite{Bowen2023CHGNet} machine learning interatomic potential.

\section{Methods}

To investigate the properties of domain boundaries in multi-domain Spinel, we construct interfaces between fully-ordered Spinel variants. The interfaces can be defined by the Spinel variants ($V_1$ and $V_2$) forming the interface and plane along which they are formed. We consider all symmetrically distinct interfaces along the (100), (110), (111), (210), (211), (221) planes of the cubic rocksalt parent phase. To construct the interfaces we fix one of the Spinel variants ($V_1$ = $\alpha_1$) and vary the other variant ($V_2$ = $\alpha_2$, $\alpha_3$,..., $\beta_4$). The variant $V_1$ is fixed to $\alpha_1$ in order to avoid redundant interfaces in which $V_1$ and $V_2$ are related to each other by the same symmetry operations. We calculate the (i) interface formation energies, (ii) Li-transport across the interfaces, (iii) possibility of 0-TM percolation across the interface and (iv) defect energy of Ti and Vacant $16d$ sites (for Li-excess compositions) in the vicinity of the interfaces.

To efficiently compute the energy and interatomic forces, we employed a fine-tuned Crystal Hamiltonian Graph Neural Network (CHGNet)\cite{Bowen2023CHGNet} as a surrogate model for DFT calculations. All DFT calculations used for fine-tuning were performed with the \texttt{VASP} package using the projector-augmented wave method \cite{Kresse1996_VASP, Kresse1999_PAW}, a plane-wave basis set with an energy cutoff of 680 eV, and a reciprocal space discretization of 25 \textit{k}-points per \AA$^{-1}$. The calculations were converged to $10^{-6}$ eV in total energy for electronic loops and 0.02 eV/\AA\ in interatomic forces for ionic loops. The regularized strongly constrained and appropriately normed meta-GGA exchange-correlation functional (r$^2$SCAN) \cite{Sun2015_SCAN, Furness2020_r2SCAN} was used with consistent computational settings as \texttt{MPScanRelaxSet} \cite{Kingsbury2022_r2scan_PRM}. The model is trained over various atomic configurations to sample the potential energy surface in the \ce{Li_{x}Mn_{0.8}Ti_{0.1}O_{1.9}F_{0.1}} chemical space. Details regarding choice of training set will be discussed in will be discussed in Ref. xxx (unpublished). The model achieves a convergence of 2 meV/atom in energy, 68 meV/\text{\AA} in force, and 0.086 GPa in stress for the test error.



The interfacial energy and machine learning molecular dynamics (MLMD) calculations are performed on cells with mutually orthogonal cell vectors. The cells contain two interfaces along the $x-y$ plane of the cell (at $z$ = 0 and 0.5). The cell sizes are chosen such that the occupancy of all Spinel variants in the cell can be distinguished from each other and the distance between the two interfaces ($d$) in the cell is at least 10 \text{\AA}. The interface cells were constructed by first assigning the occupancy of $V_1$ and $V_2$ in the lower ($0 < z \le$ 0.5) and upper half ($z >$ 0.5) of the cell on a rock-salt lattice respectively and subsequently performing geometric relaxation on the structure. The interfacial energies are calculated at the fully delithiated composition of \ce{Mn2O4}. The MLMD simulations were performed under the constant number of particles, volume, and temperature (NVT) ensemble, with a time step of 2 fs at $T = 1100 K$ and for the \ce{Li_{0.25}Mn2O4} composition. In order to study the effect TM ordering at the interface on the Li-transport the position of Mn atoms within the cell was fixed. The time for which the MLMD simulations were performed was determined by monitoring the mean-squared-displacement (MSD) along the direction perpendicular to the interface ($z$-direction). The MLMD was run long enough such that for MSD$_z$ $\ge \frac{d}{2}^2$ to confirm percolation of lithium through the interface or till a plateau (MSD $\propto t^0$) is observed in MSD$_z$ vs $t$ (in case of a non-percolating interface). The mean-squared-displacement was calculated using full time-window averaging.\cite{Giorgino2019_JOSS_2019_MSD,YifeiMo_2018_MD_Details_Paper_npj_CompMater}


\section{Results}

\begin{figure}[hbt!]
\centering
\includegraphics[scale=0.3]{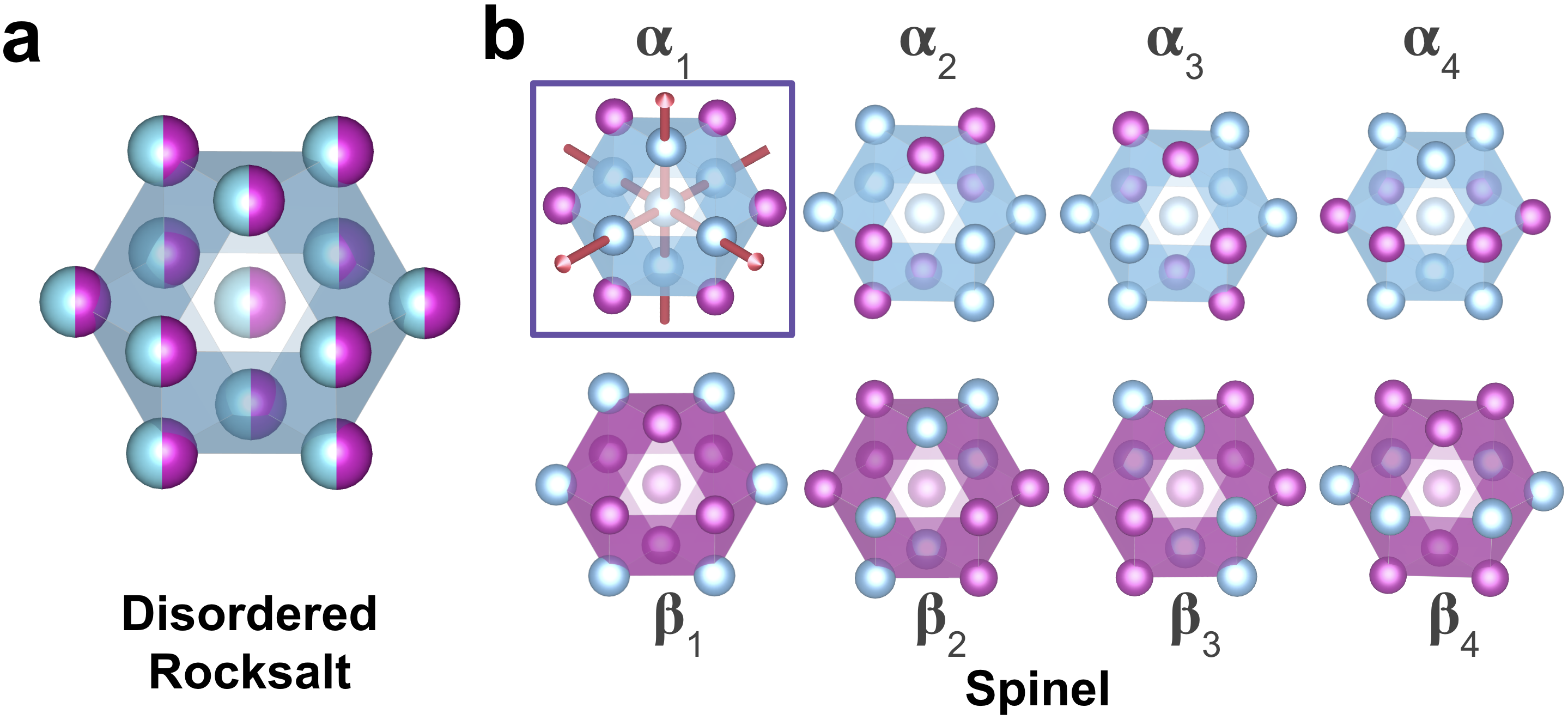}
\caption{(a) Polyhedral motif of octahedral cation sites in the cubic disordered rocksalt (DRX) structure at \ce{Li2Mn2O2} composition viewed down the \textless111\textgreater axes. The motif consists of 13 sites, which includes a central site and its 12 nearest neighbor octahedral sites, shown with partial occupancies of Li (light blue) and Mn (purple). No tetrahedral sites are shown. (b) The eight distinct crystallographic orientations of the Spinel structure in the same motif as shown in (a). The choice of $16c$ and $16d$ sub-lattices is respectively shown for each spinel variant by light blue and purple. Spinel orientations related to each other through complete exchange of the definition of the $16c$ and $16d$ sub-lattices are named $\alpha_i$ and $\beta_i$, where $i=1,2,3,4$ represent orientations which are related to each other with a 2-fold rotation along the \textless110\textgreater axes. The 2-fold axes of rotations in $\alpha_1$ relating it to other $\alpha_i$ are shown by red arrows.}
\label{fgr:Crystallography}
\end{figure}


Phase transformation of a high symmetry parent lattice into a lower symmetry superstructure results in multiple ways in which the superstructure can be arranged on the parent lattice. These variants of the superstructure can therefore be distinguished from each other based on their occupancies of the parent lattice. DRX belongs to the $Fm\bar{3}m$ space group, which is the rocksalt structure (RS) whereas the $\delta$-phase belongs to $Fd\bar{3}m$ space group.\cite{HanMingHau2024earth,Tucker_2024_Pulsing_AM} Using group theory analysis one can derive rules which dictate the number of variants during such a transformation, which is the number of elements in the quotient group of $Fm\bar{3}m$ divided by $Fd\bar{3}m$ \cite{wondratschek1974crystallographic, norrby1990factor}. The phase transformations result in the formation of 8 cosets which can be represented using Seitz notation as: ($\mathbb{I}$ $\vert$ 0\ 0\ 0),  ($\mathbb{I}$ $\vert$ 0\ 1/4\ 1/4), ($\mathbb{I}$ $\vert$ 0\ 0\ 1/2),  ($\mathbb{I}$ $\vert$ 1/4\ 1/2\ 1/4), ($\mathbb{I}$ $\vert$ 0\ 1/4\ 3/4),  ($\mathbb{I}$ $\vert$ 3/4\ 1/2\ 3/4),  ($\mathbb{I}$ $\vert$ 3/4\ 3/4\ 0), and ($\mathbb{I}$ $\vert$ 1/4\ 1/2\ 3/4) where $\mathbb{I}$ is the identity operation.\cite{yasuda2020radiation} Simply put, these 8 cosets represent the translational symmetry operations which relate a particular Spinel variant to itself (identity operation) and the other Spinel variants. 



To visualize the Spinel variants on the parent rocksalt lattice more intuitively, Figure \ref{fgr:Crystallography} shows the occupancy of DRX and the different Spinel variants on a structural motif with 13 octahedral sites for the typical rock-salt \ce{LiMnO2} and Spinel \ce{LiMn2O4} stoichiometries. Since the octahedral occupancy is sufficient to distinguish the Spinel variants, the tetrahedral sites are not shown. In the DRX motif, all the octahedral sites have the same partial occupancy of 50\% each for Li (light blue) and Mn (purple), making the octahedral sites indistinguishable from one another. In the cubic \ce{LiMn2O4} Spinel phase, the Li enters the tetrahedral sites and the octahedral sites --- described as $16c$ and $16d$ sites in the Wyckoff notation --- can be distinguished from each other based on whether they are vacant ($16c$, light blue) or occupied by Mn ($16d$, purple). The motif of a particular Spinel variant can be related to the others through either (i) a simple exchange of the $16c$ and $16d$ sub-lattices or (ii) applying a rotational (three 2-fold 110 axes) symmetry operation or (iii) applying a combination of (i) and (ii). The Spinel variants related to each other through an exchange of the $16c$ and $16d$ sub-lattices are named $\alpha_i$ and $\beta_i$ ($i=1,2,3,4$) and are shown in the same column in Figure \ref{fgr:Crystallography}. The subscript $i$ in $\alpha_i$/$\beta_i$ represent variants which are rotationally related to each other and are shown in the same row in Figure \ref{fgr:Crystallography}. Examples of the 2-fold axes of rotation relating $\alpha_1$ (purple box) to the other $\alpha_i$ are drawn in $\alpha_1$ using red arrows. 

\begin{figure}[hbt!]
\centering
\includegraphics[scale=0.23]{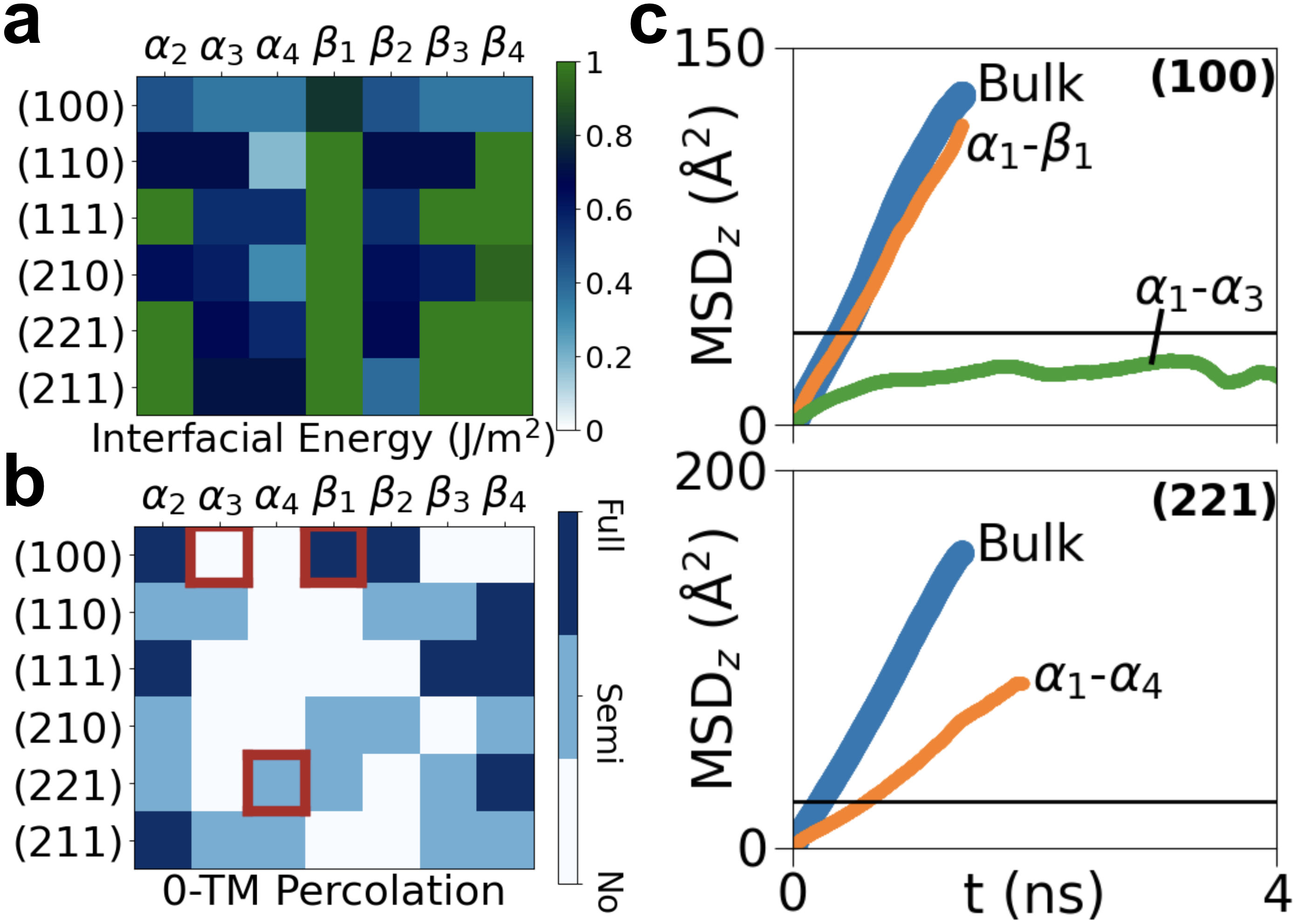}
\caption{(a) Interfacial Energy of the interfaces between $\alpha_1$ and the 7 other spinel variants along six different low-index planes. (b) Percolation of 0-TM channels across the interfaces shown in (a). Interfaces preserving the total number of 0-TM channels present in pristine spinel are fully percolating and are shown in dark blue. Non-percolating and semi-percolating interfaces are shown in white and light blue respectively. (c) Li mean-squared-displacement in the direction perpendicular to the interface (MSD$_z$) versus $t$ as obtained from MLMD simulations for the three interfaces outlined in red in panel (b). The black line corresponds to $MSD_z = \frac{d}{2}^2$, where $d$ is the distance between two interfaces within the simulation cell.}
\label{fgr:Interface_Properties}
\end{figure}

The DRX compositions for which the transformation to the $\delta$-phase is observed, have some octahedral Mn in the \ce{LiMnO2} stoichiometry substituted with small amounts of Li (the so-called Li-excess compositions) and Ti (e.g. \ce{Li_{1.05}Mn_{0.85}Ti_{0.1}O_2}).\cite{Zijian_2024_Nature_Energy} However, to understand the type of boundaries forming between the domains we first study interfaces between Spinel variants without any Ti or excess-Li (Mn occupies all $16d$ sites).\cite{Zijian_2024_Nature_Energy} Figure \ref{fgr:Interface_Properties}a shows the heat map for calculated interfacial energies ($\gamma$) between variants. The numerical values of the interfacial energies are also provided in the table \ref{tbl:Interfacial_Energies}. The rows in the heat map correspond to the six low-index planes along which the interfaces are constructed. The columns in the heat map correspond to the variant (named at the top of the column) that is in contact with $V_1$. To distinguish the energies of the lower energy interfaces, the colorbar for the heatmap is saturated at $\gamma =$ 1 J/m$^2$. We find that for all instances of $V_2$, the interface formed along the (100) planes are either the lowest or close to the lowest in energy. For $V_2$=$\alpha_4$ and $V_2$=$\beta_2$, the lowest energy interfaces form along (110) and (211) planes respectively. In general, the highest energy interfaces are formed between the variants $\alpha_1$ and $\beta_1$ for which the $16c$ and $16d$ sub-lattices exchange completely.

\begin{table}[h!]
    \centering
    \begin{tabular}{|c||p{1cm}|p{1cm}|p{1cm}|p{1cm}|p{1cm}|p{1cm}|p{1cm}|}
        \hline
        Interface Index & $\alpha_2$ & $\alpha_2$ & $\alpha_2$ & $\beta_1$ & $\beta_2$ & $\beta_3$ & $\beta_4$ \\ 
        \hline
        (100) & 0.45 & 0.35 & 0.35 & 0.81 & 0.45 & 0.35 & 0.35 \\
        \hline
        (110) & 0.7 & 0.7 & 0.18 & 1.48 & 0.7 & 0.7 & 1.39 \\ 
        \hline
        (111) & 2.06 & 0.55 & 0.55 & 2.79 & 0.55 & 2.06 & 2.06 \\ 
        \hline
        (210) & 0.64 & 0.59 & 0.3 & 1.18 & 0.64 & 0.59 & 0.94\\ 
        \hline
        (221) & 1.3 & 0.65 & 0.57 & 1.39 & 0.65 & 1.3 & 1.49 \\ 
        \hline
        (211) & 1.78 & 0.72 & 0.72 & 2.2 & 0.38 & 1.41 & 1.41 \\ 
        \hline
    \end{tabular}
    \caption{Interfacial Energies (J/m$^2$) of the interfaces between $\alpha_1$ and the 7 other spinel variants along six different low-index planes.}
    \label{tbl:Interfacial_Energies}
\end{table}

The formation of interfaces between Spinel variants can reduce the number of 0-TM channels near the interface in comparison to the fully ordered Spinel,\cite{Urban2014_AEM} thereby impacting the Li-transport. Figure \ref{fgr:Interface_Properties}b shows the heatmap for 0-TM percolation across the interfaces shown in Figure \ref{fgr:Interface_Properties}a. Interfaces which preserve the number of 0-TM channels found in the pristine Spinel are fully percolating and are shown with dark blue. On the other hand, interfaces with no 0-TM pathways available for Li-percolation are shown in white. Semi-percolating interfaces with fewer 0-TM passages compared to pristine Spinel are shown in light blue. We find that the majority of the interfaces are either non-percolating or semi-percolating. Among the lowest energy interfaces between the seven variants only two cases ($\alpha_1$-$\alpha_2$ and $\alpha_1$-$\beta_1$) show 0-TM percolation.


In addition to 0-TM percolation, the ordering of cations around the 0-TM channels has also been shown to influence transport by varying the Li site-energetics\cite{Anand2023PRM_DRX}. So even in cases when the interface has the same number of 0-TM pathways as fully-ordered Spinel, the change in cation ordering around the interface can play a role in Li-transport. To further investigate Li transport through the various interfaces we performed MLMD simulations at $T = 1100 K$ in cells containing two interfaces along the $x-y$ plane of the cell with the distance between the two interfaces ($d$) at least 10 \text{\AA}. Figure \ref{fgr:Interface_Properties}c shows Li mean-squared displacement (MSD) versus $t$ in the direction perpendicular to the interface (\textit{z}-direction of the simulation cell). The results are shown for three interfaces which include a fully-percolating ($\alpha_1$-$\beta_1$ along 100), a non-percolating ($\alpha_1$-$\alpha_3$ along 100) and a semi-percolating ($\alpha_1$-$\alpha_4$ along 221) example as indicated by red squares in Figure \ref{fgr:Interface_Properties}b. The corresponding MSD for the bulk Spinel is also displayed for comparison. The fully percolating interface shows the same Li-transport as the bulk Spinel. The MSD in the non-percolating interface on the other hand plateaus at MSD$<$($\frac{d}{2}$)$^2$ for $t > $ 300 ps, signifying negligible Li-transport across the interface. The semi-percolating $\alpha_1$-$\alpha_4$ interface along (221) shows only partial Li transport, with a Li diffusivity smaller than bulk Spinel by a factor of $\sim 2.5$. These results are consistent with the 0-TM percolation analysis (see Figure \ref{fgr:Interface_Properties}b) and confirm the importance of 0-TM channel availability for li-transport.

\begin{figure}[hbt!]
\centering
\includegraphics[scale=0.18]{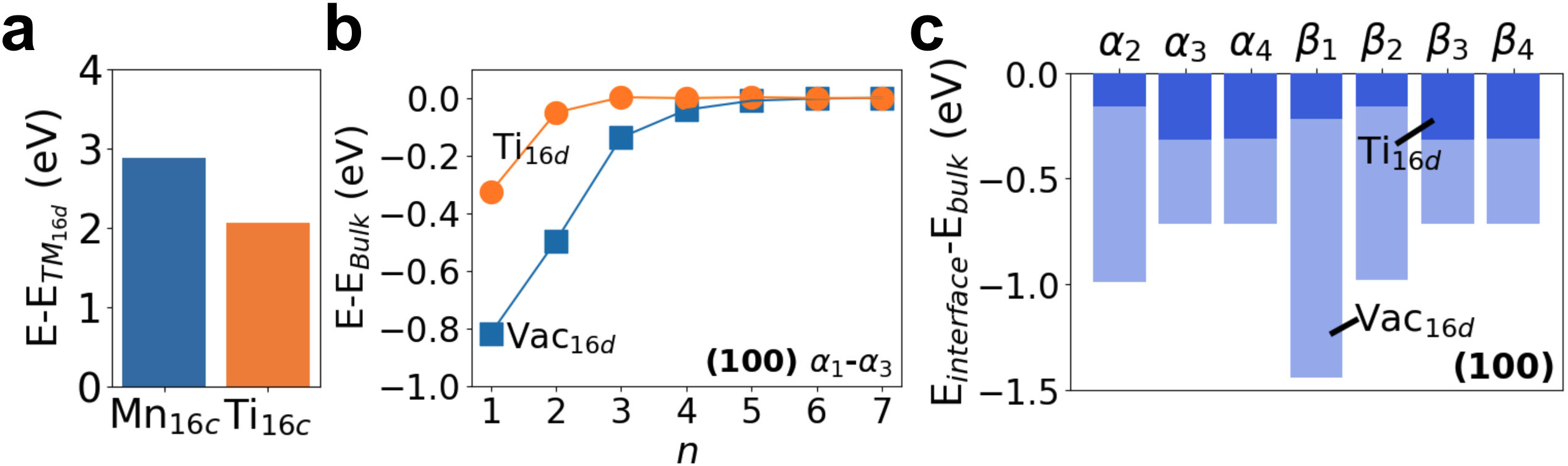}
\caption{(a) Energy of defect structures containing Ti and Mn atoms placed in the $16c$ (anti-site defect) Wyckoff sites with respect to bulk Spinel ordered structure where all cations are placed on $16d$ sites. (b) Energy change for Ti$_{16d}$ and Vac$_{16d}$ defects upon segregation of the defect from the bulk Spinel to the (100) interface between $\alpha_1$ and $\alpha_3$. The distance of the defect from the interface is given by the number of atomic layers ($n$), $n=1$ represents the interface. (c) Segregation energy of Ti$_{16d}$ and Vac$_{16d}$ defects at different (100) interfaces. (d) Schematic showing two competing scenarios for interface formation between Spinel domains when the Ti coordinates are fixed. }
\label{fgr:Defects_Competition}
\end{figure}

To study the effect of changes to the structure arising from adding Ti and excess-Li to the composition, we consider the defect energy of Ti and $16d$ vacancies (Vac$_{16d}$) in bulk and at the Spinel interfaces. We note that Vac$_{16d}$ in the Spinel structure is evaluated here because they form when Li-excess compositions --- in which Li substitutes for Mn as a cation --- are delithiated (e.g. \ce{Li_{1.05}Mn_{0.85}Ti_{0.1}O_2} $\rightarrow$ \ce{Li_{0.45}Mn_{0.85}Ti_{0.1}O_2}). Since most of the DRX-to-$\delta$-phase transformation occurs at delithiated compositions, it is important to study where Vac$_{16d}$ preferentially reside within the Spinel domains. Figure \ref{fgr:Defects_Competition}a compares the Spinel anti-site (single cation placed at the vacant $16c$ site instead of the cation $16d$ site of the Spinel structure) formation energy of Mn and Ti. Similar to Mn, Ti has a large anti-site defect energy, meaning that Ti strongly favors Spinel ordering upon delithiation. The larger anti-site defect energy for Mn compared to Ti suggests that it has the stronger driving force to form the Spinel ordering by occupying the $16d$ sites. Figure \ref{fgr:Defects_Competition}b shows the change in energy when the Ti$_{16d}$ and Vac$_{16d}$ defect segregates from the bulk Spinel to the (100) interface $\alpha_1$-$\alpha_3$. The distance of the defect from the interface is given by the number of atomic layers $n$ from the interface, where $n=1$ represents the atomic layer at the interface. We find that the energy of both defects drop sharply as the defects approach within 2 atomic layers of the interface. A comparison of the segregation energy ($E_{interface}-E_{bulk}$) for both the Ti$_{16d}$ and Vac$_{16d}$ defects at all the low energy (100) interfaces (see Figure \ref{fgr:Defects_Competition}c) reveals the same affinity for the interface in all cases. In general, $16d$ vacancies have a much stronger driving force to segregate to the interface than Ti.  

\begin{figure}[hbt!]
\centering
\includegraphics[scale=0.28]{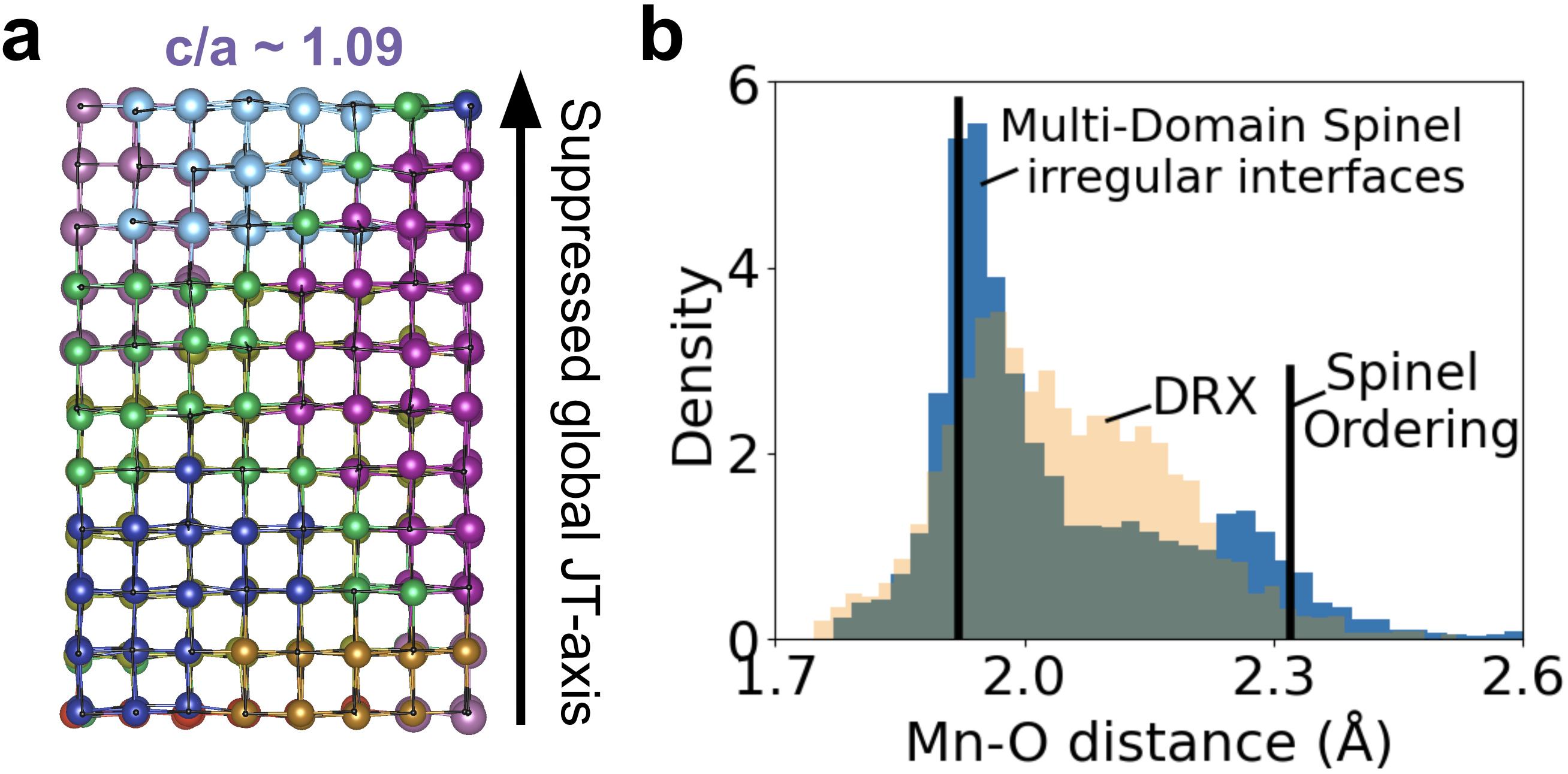}
\caption{(a) Region of \ce{LiMnO2} structure containing multiple nanoscale domains of Spinel-type ordering corresponding to the \ce{Li_{0.5}MnO2} composition. The different domains can be distinguished by the color of the cation sites. The oxygen atoms are not shown. The interfaces between domains are irregular and do not correspond to the lowest energy interfaces. (b) Histogram of Mn-O distances in \ce{Li_{1.05}Mn_{0.85}Ti_{0.1}O2} in DRX configuration and multi-domain Spinel configuration with irregular interfaces as shown in (a). The Mn-O distances in Jahn-Teller distorted tetragonal \ce{LiMnO2} with Spinel Li-Mn ordering are shown by black colored bars.}
\label{fgr:C_to_T_Suppresion}
\end{figure}

When bulk \ce{LiMn2O4} is lithiated it undergoes a collective two-phase cubic-tetragonal transformation as the tetrahedral $8a$ occupancy of Li has to change to complete $16d$ occupancy. Because of the reduction of all Mn to $3+$ in the \ce{Li2Mn2O4} end-member phase a collective Jahn-Teller (JT) distortion also occurs. The large strain caused by the inhomogeneity of the transformation degrades the material; making it unusable in the 3V region. This transformation has been found to be suppressed in the multi-domain $\delta$-phase which shows solid-solution behavior in the 3V region.\cite{Zijian_2024_Nature_Energy} To isolate the effect that the domain boundaries have on the structure and the JT distortion of the lithiated tetragonal Spinel from that of cation disorder, we simulate a multi-domain structure with no TM-disorder inside the domains or at the boundaries between them. We deliberately construct domains with irregular domain morphologies and non-planar boundaries, which will be discussed below is a consequence of introducing Ti in the $\delta$-phase composition. The domain size is approximately 1.6 nm across all domains. Initially, we consider only the \ce{Li2Mn2O4} composition for comparison with the fully-ordered single-domain structure. Figure \ref{fgr:C_to_T_Suppresion} a shows a region of the multi-domain \ce{LiMnO2} structure where many domains form interfaces between each other. The cation species (Li and Mn) within each domain are given the same color in order to easily distinguish one domain from another. We find that even though the system has a multi-domain structure its JT-axis is aligned across the domain boundaries (see Figure \ref{fgr:C_to_T_Suppresion} a) forming a collective JT distortion. However, the c/a ratio of the multi-domain structure is significantly lower (1.09) than that of a single domain tetragonal bulk Spinel structure (1.2). 

Figure \ref{fgr:C_to_T_Suppresion}b compares the distribution of Mn$-$O distances in the same multi-domain Spinel structure (blue) with that of the DRX structure (orange) at the \ce{Li_{1.05}Mn_{0.85}Ti_{0.1}O_{2}} composition. The peaks corresponding to Mn$-$O distances for single domain tetragonal bulk \ce{LiMnO2} Spinel structure are shown in black. The multi-domain spinel structure was obtained by randomly substituting Mn in the Figure \ref{fgr:C_to_T_Suppresion}a structure with Ti and Li. This change in composition from \ce{Li2Mn2O4} to \ce{Li_{1.05}Mn_{0.85}Ti_{0.1}O2} has a negligible impact on the c/a ratio of the multi-domain spinel structure. We find that the multi-domain Spinel structure --- similar to the DRX structure at the same composition --- has a distribution of nearest neighbor Mn-O distances (see Figure \ref{fgr:C_to_T_Suppresion}b). Unlike the DRX structure however, the multi-domain Spinel structure has a bimodal Mn-O distribution with peaks lying close to the tetragonal \ce{LiMnO2} structure without any domain boundaries (vertical black lines in Figure \ref{fgr:C_to_T_Suppresion} b). The peaks in the distribution for the multi-domain structure are closer to each other compared to the single domain tetragonal bulk Spinel structure, indicative of a suppressed JT-distortion.

\section{Discussion}

Microscopy images of the transformed spinel-like nanostructures in previous work often show regions where the domains form tri-junctions between each other.\cite{YMChiang_2000_Layered_to_Delta_ChemMater} This result can only be explained if the occupancy of three mutually adjacent domains belong to different variants of Spinel, meaning that at least three Spinel variants must exist, consistent with the group theory arguement made in this paper that eight possible variants exist on the rocksalt lattice. Our calculations reveal that the interfaces between variants related through an exchange of the $16c$ and $16d$ sub-lattices are generally the highest in energy. The higher energy of these interfaces can be attributed to the complete mismatch in occupancies between the variants at the interface. In the case of other interfaces, which are lower in energy, this mismatch in occupancy at the interface is only partial.



Our results show that for compositions without any Li-excess, most Spinel interfaces show quite poor Li-percolation (see Figure \ref{fgr:Interface_Properties}). Over three-quarters of the low index Spinel interfaces considered here either show partial 0-TM percolation or no percolation at all. The lowest energy interfaces of a Spinel variant with five other Spinel variants (out of seven) are non-percolating (see Figure \ref{fgr:Interface_Properties}). Furthermore, even these five non-percolating interfaces of $\alpha_1$ will be favored over the percolating ones due to their lower energies. These results are consistent with the fact that while $\delta$-phase has higher rate capability than fully-disordered DRX \cite{Zijian_2024_Nature_Energy} its rate performance is likely still below that of bulk Spinel. 

\begin{figure}[hbt!]
\centering
\includegraphics[scale=0.25]{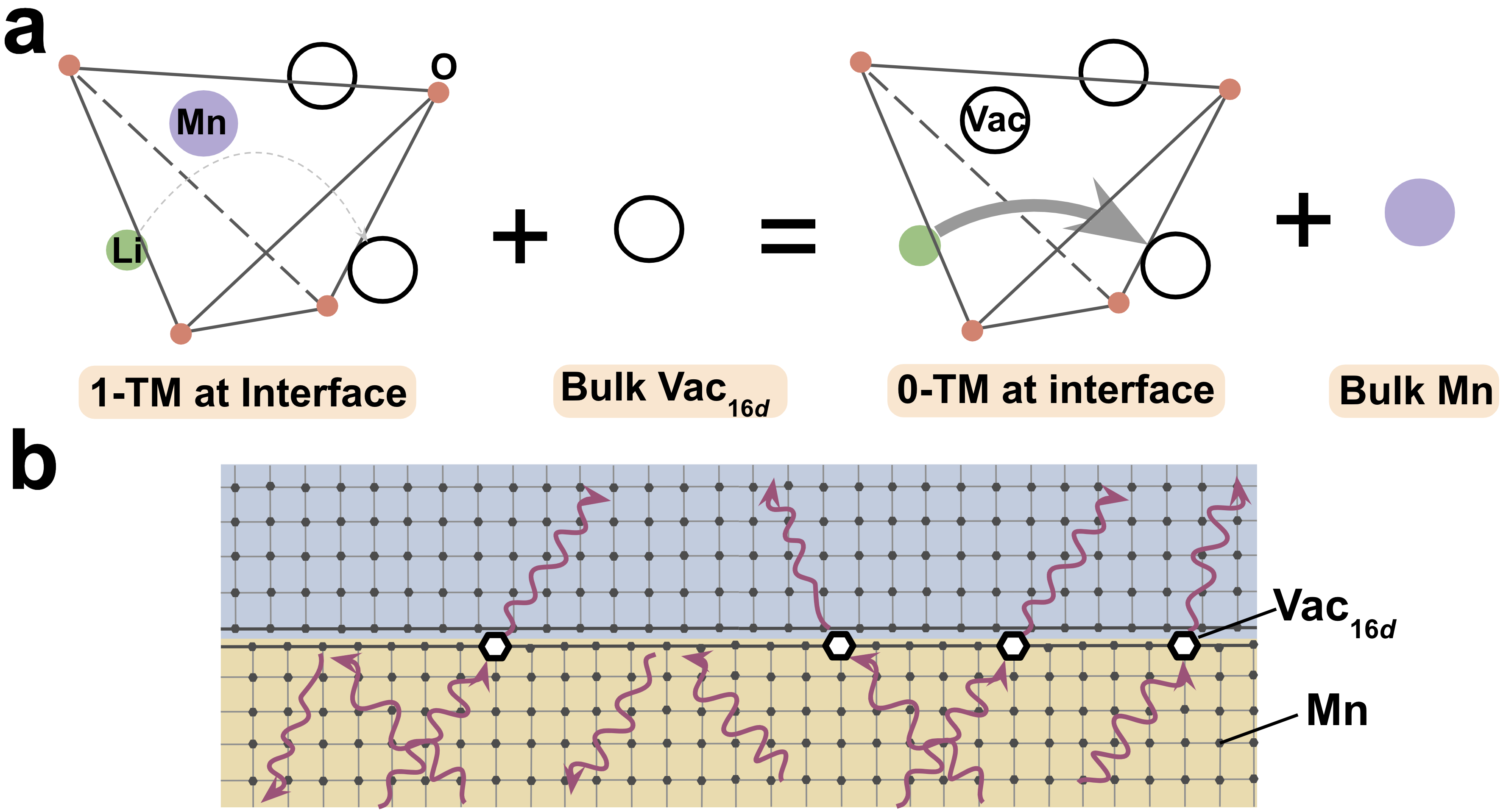}
\caption{(a) Reaction describing how 1-TM environment at the interface between Spinel variants can be modified to a 0-TM environment if the octahedral Mn face-sharing with the 1-TM tetrahedral pathway migrates to the bulk of the Spinel domain. (b) Sketch of an interface between Spinel variants which is 0-TM non-percolating at the conventional (Li/Vac):TM=1:1 cation stoichiometry but becomes 0-TM percolating due to the segregation of Vac$_{16d}$ defects (large empty hexagons) at the interface. The squiggly maroon arrows represent pathways taken by Li.}
\label{fig:Interfacial_1TM_to_0TM}
\end{figure}

\begin{figure}[hbt!]
\centering
\includegraphics[scale=0.25]{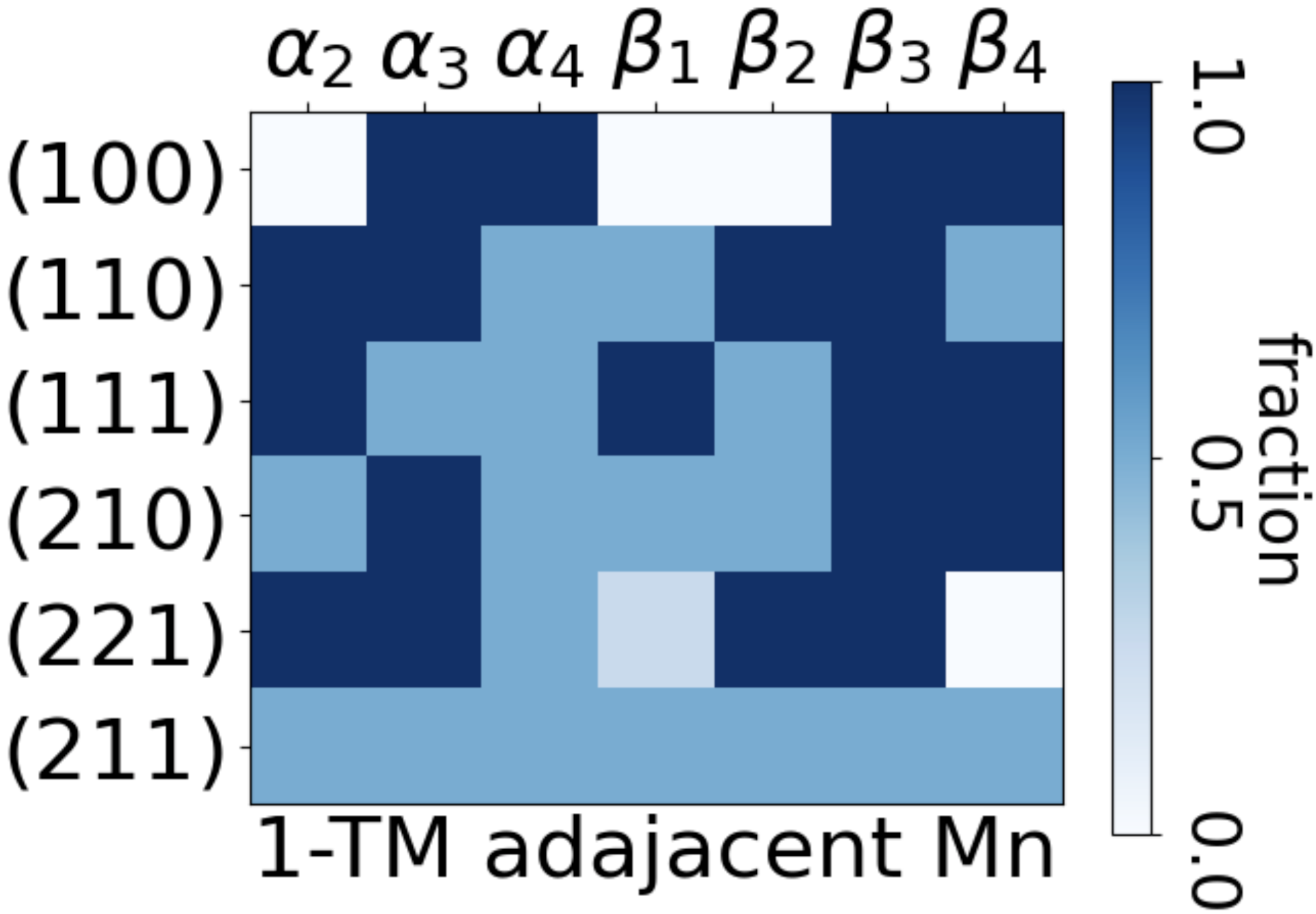}
\caption{Fraction of Mn at various interfaces which are a part of a tetrahedral 1-TM environments. The same interfaces as in Figure \ref{fgr:Interface_Properties} are considered.}
\label{fig:Interfacial_1TM_Mn}
\end{figure}

The long-range Li-percolation in $\delta$-phase may however benefit from the Li-excess compositions in the DRX structure. In the Li-excess composition, the Spinel-like domains in $\delta$-phase will contain $16d$ sites which are not occupied by TM. Our results in Figure \ref{fgr:Defects_Competition} c show that when Li is removed during electrochemical cycling, the vacant $16d$ sites (Vac$_{16d}$) have a strong preference to be a part of the interfaces (see Figure \ref{fgr:Defects_Competition} c). This lack of TM at the interface can be beneficial in enhancing the rate capability of the $\delta$-phase if these vacancies opens up new 0-TM pathways. The reaction in Figure \ref{fig:Interfacial_1TM_to_0TM}a describes how the migration of a Mn which is a part of a 1-TM tetrahedral environment at the interface to a 16$d$ site in the bulk of the domain leaves behind a Vac$_{16d}$ and open up new 0-TM pathway. As a result, even an interface which was otherwise 0-TM non-percolating at the conventional (Li/Vac):TM=1:1 cation stoichiometry can allow long-range 0-TM Li-percolation if Vac$_{16d}$ segregates to a $16d$ site at the interface which was previously occupied by a Mn that formed a 1-TM environment. Figure \ref{fig:Interfacial_1TM_to_0TM} b sketches Li-transport pathways (shown in maroon arrows) near one such interface; the Li atoms are only able to pass through regions in interface where Vac$_{16d}$ defects have segregated. 


It is therefore crucial to identify which Spinel interfaces create 1-TM tetrahedral channels and how many of the Mn at the interface are a part of 1-TM environments. Figure \ref{fig:Interfacial_1TM_Mn} shows a heatmap for the fraction $f$ 

\[
f = \frac{\text{number of Mn at interface which are a part of 1-TM environments}}{\text{total number of Mn at interface}
}
\]

This structural analysis is performed at the conventional (Li/Vac):TM=1:1 cation stoichiometry. We note that almost all interfaces contain some Mn which are involved in formation of 1-TM pathways, suggesting that most interfaces --- unlike fully-ordered bulk Spinel --- contain 1-TM environments. All interfaces that are 0-TM non-percolating and semi-percolating (see Figure \ref{fgr:Interface_Properties} b) contain 1-TM pathways, meaning that the a segregation of Vac$_{16d}$ to these interface could open up a new 0-TM sites. In all the low-energy 0-TM non-percolating (100) interfaces the transition metal sites at the interface \textit{always} lie next to a 1-TM tetrahedral site, meaning that the preferential segregation of Vac$_{16d}$ to these interfacial TM as shown in Figure \ref{fgr:Defects_Competition}b, would automatically ensures 0-TM percolation across the interface (see Figure \ref{fig:Interfacial_1TM_Mn}). Nearly half (16 out of 33) of the non-percolating and semi-percolating interfaces studied here have the same structural feature where substitution of \textit{any} Mn with a vacancy/Li at the interface guarantees opening up new 0-TM pathways (see Figure \ref{fig:Interfacial_1TM_Mn}). Since the DRX-to-$\delta$-phase transformation occurs through vacancy mediated migration of Mn, the vacant $16d$ sites can be expected to rearrange throughout the transformation and move with the domain boundaries during changes in the nanostructure. Li-excess is therefore crucial for enhancing Li-transport in the $\delta$-phase at all stages of the transformation.


The large anti-site energy of Ti, suggests that it also strongly favors the $16d$ site in Spinel ordering just like Mn. This result is in line with experimental observations where the spinel-like domains of the $\delta$-phase are largely ordered through the transformation even at very small domain sizes.\cite{Tucker_2024_Pulsing_AM} In addition to the stronger thermodynamic driving force for Mn to form Spinel compared to Ti,\cite{Gerd_Anton_1999_LTMO_EChemActa} previous work\cite{Zijian_2024_Nature_Energy} has also shown Mn to have smaller migration barriers compared to Ti. Since both thermodynamic and kinetic factors point towards the higher mobility of Mn relative to Ti, in the initial parts of the transformation Mn can be expected to move around the relatively immobile Ti. This Mn rearrangement will occur to ensure that both the immobile Ti as well as the migrated Mn occupy their favored Spinel $16d$ sub-lattice positions. As a result, the initial occupancy of Ti in the DRX phase can influence the local Spinel ordering of Mn around it significantly. The random distribution of Ti in the initial DRX structure will ensure that not all Ti can occupy sites of the same Spinel variant, contributing to the multi-domain nanostructure observed in the $\delta$-phase. 


The extent to which the domains can grow will then depend on the length-scale over which Ti occupancy in the the initial DRX structure is consistent with a particular Spinel variant. In this model, a spinel domain will form around Ti by the Mn ions rearranging around it to form a short-coherence length occupied $16d$ sub-lattice which grows in all possible directions untill Ti in the ``wrong" sites are encountered. This Ti-dominated mechanism makes the length scale dependent on the Ti-composition of the sample. Furthermore, the affinity of Ti to occupy the boundary (see Figure \ref{fgr:Defects_Competition}c) can prevent the boundaries from moving during the transformation. This pinning effect from Ti can prevent coarsening and domain growth in the $\delta$-phase. These effects explain why the domain size starts to saturate in the $\delta$-phase when it is obtained in electrochemical cycling\cite{Tucker_2024_Pulsing_AM} or by low temperature annealing.\cite{HanMingHau2024earth}

\begin{figure}[hbt!]
\centering
\includegraphics[scale=0.26]{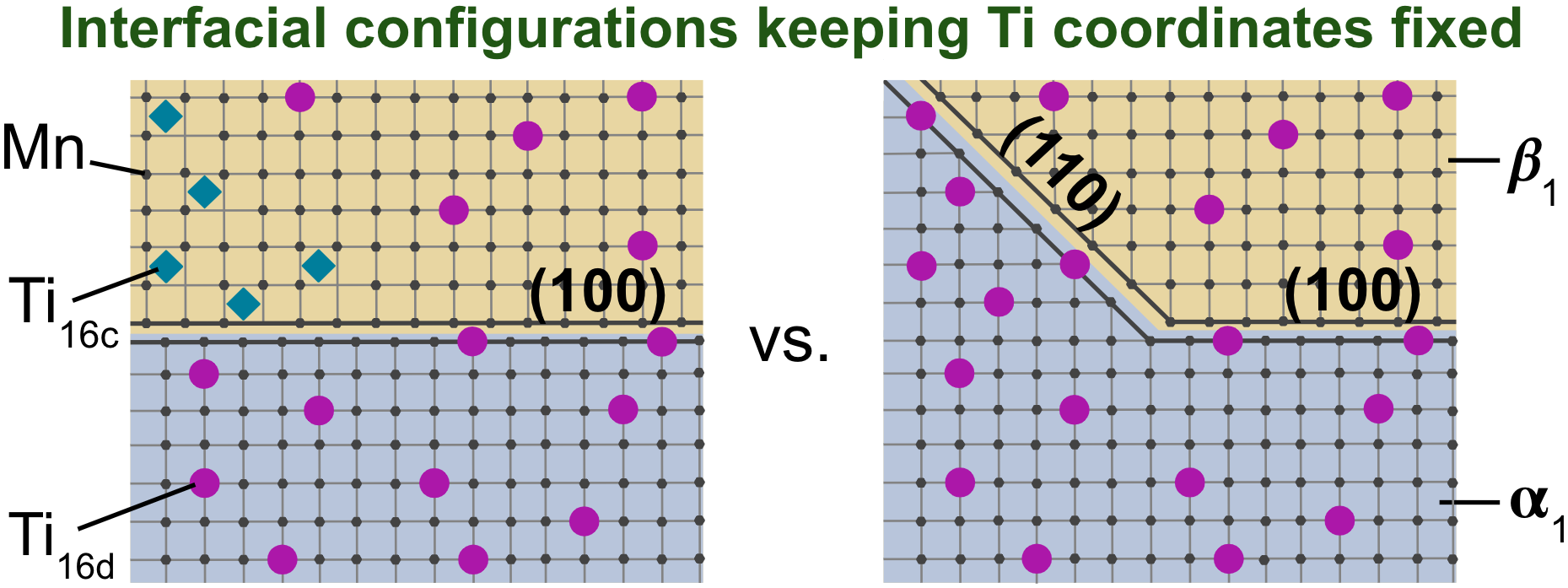}
\caption{Schematic showing two competing scenarios for interface formation between between Spinel domains when the Ti coordinates are fixed. The left figure shows a domain with some Ti anti-sites but a low energy interface plane, whereas the right figure has all Ti on 16$d$ sites at the cost of forming a higher energy interfaces.}
\label{fgr:Competing_Interfacial_Senarios}
\end{figure}

To accommodate the Ti occupancies from the initial disordered structure into the $16d$ sub-lattices of a Spinel variant the domains may take on irregular shapes with higher energy interfaces between domains instead of only the lower energy interfaces. Figure \ref{fgr:Competing_Interfacial_Senarios} sketches the competition between correct Ti occupancy and low energy interfaces leading to two hypothetical scenarios of anti-phase boundary formation between Spinel variants. The Ti coordinates are kept fixed in both scenarios and only the Mn positions (small hexagons) are allowed to change. The $16d$ sub-lattice in both variants $\alpha_1$ and $\beta_1$ is connected by a mesh of lines and the square boxes formed by this mesh indicate the $16c$ site positions. The Spinel orientation of the two domains forming the boundary are distinguished by the color shade in the background. In the first case, Ti occupies only the $16d$ sites in $\alpha_1$ domain but occupies a few higher energy $16c$ sites in $\beta_1$ but the interface remains limited to the low energy (100). In the second case, the Mn atoms rearrange to accommodate all Ti on either side of the boundary on the energetically favored $16d$ sites by creating a higher energy (110) interface and increasing the size of the $\alpha_1$ domain. 

The second scenario in Figure \ref{fgr:Competing_Interfacial_Senarios} is an example of a domain boundary that deviates from its low-energy orientation to accommodate Ti on $16d$ positions. For the formation of such irregular boundaries, the energy penalty associated with the formation of the higher energy interfaces ($\Delta E_{\text{Interface}} = E_{\text{int-high}} \sigma_{\text{int-high}} - E^{\text{int-low}} \sigma_{\text{int-low}}$) must be overcome by the energy gain associated with Ti placement on the $16d$ sub-lattice ($E_{\text{Ti-ordering}}$). The energy gain from occupancy of all Ti on the Spinel $16d$ sites will be largely determined by the Ti composition itself $E_{\text{Ti-ordering}} = \rho_{\text{Ti}} \times E_{\text{Ti}}^{anti-site}$, where $\rho_{\text{Ti}}$ is the density of Ti-species in the structure and $E_{\text{Ti}}^{anti-site}$ ($\sim$ 2 eV/ Ti is the Ti anti-site energy (Figure \ref{fgr:Defects_Competition}a). For the \ce{Li_{1.05}Mn_{0.85}Ti_{0.1}O2} composition, $\rho_{\text{Ti}}\sim$3 Ti/ nm$^3$ and $E_{\text{Ti-ordering}} \sim$ 6 eV/nm$^3$. On the other hand, for a fixed interfacial area the energy penalty of forming an alternate higher energy interface between two Spinel variants will depend on its energy difference with respect to the lowest energy interface $\Delta E_{\text{Interface}}|_{\sigma} = E_{\text{int-high}} - E^{\text{int-low}}$. We note that $\Delta E_{\text{Interface}}|_{\sigma}$ for over half of the higher energy interfaces lie in the range 0.4 eV/nm$^2-$2.5 eV/nm$^2$ (supplementary figure \ref{fig:Supp4}). In the case of these interfaces, occupancy of all Ti species on $16d$ sites will allow for 2.4-15 nm$^2$ ($\frac{E_{\text{Ti-ordering}}}{\Delta E_{\text{Interface}}|_{\sigma}}$) area of higher energy interfaces to form within 1 nm$^3$ volume of $\delta$-phase. As a result, the initial occupancy of the relatively immobile Ti in the $\delta$-phase composition can influence the domain morphology through the formation of irregular domain boundaries. 


An important feature of the detrimental two-phase reaction which is associated with poor cycling stability in Mn-based Spinel structure is the change in structure from cubic to tetragonal. Whereas the tetragonal JT distortion is fundamental to octahedral \ce{Mn^{3+}}, the two-phase reaction is independent of the distortion and arises from the spinel structure itself; occurring even between the two cubic phases of stoichiometric \ce{LiCo2O4} and ``low-temperature" \ce{LiCoO2}.\cite{lee2019lithiated} Nevertheless, previous work demonstrating suppression of the two-phase reaction of lithiated Mn-based Spinel also observes a suppression in tetragonal distortion along with it; suggesting a possible corelation between the two effects.\cite{ji2020_Nat_Energy_PDS,Zijian2021_Matter_ContinuousDisorder} The common approach among these studies is the use of cation disorder.\cite{Tina_2022_ACS_Energy_Lett_Two_Phase_Transition} Structural characterization of $\delta$-phase in the discharged state (towards the end of the 3 V plateau) also suggests a suppressed two-phase reaction with a solid solution lithiation of a cubic phase. However, the refinement of the XRD spectrum for the $\delta$-phase suggests Spinel domains with very low disorder levels ($<$5\%).\cite{Tucker_2024_Pulsing_AM} As such, disorder alone cannot be the cause of solid solution lithiation observed during \textit{in-situ} diffraction measurements.\cite{Zijian_2024_Nature_Energy,HanMingHau2024earth} Our results show that even highly-ordered domains with irregular domain morphologies --- which in $\delta$-phase results from the sluggish migration of randomly distributed Ti --- could suppress the tetragonal distortion upon lithiation significantly. By limiting domain growth to only a few nm, the JT distortion can be suppressed due to the strain between adjacent distorting domains, even when the interior of domains consists of fully-ordered spinel. While earlier research on transformed spinel-like structures attribute suppressed c/a ratios to the formation of ferroelastic\cite{YMChiang1999_ECSSLett_Origin_of_Transformation,YMChiang_2000_Layered_to_Delta_ChemMater} and incoherent domains where the JT-axis of different domains don't necessarily align with each other, we show that the suppression of the distortion could be observed even if the JT-axis aligns across domains. 

While just suppressing the JT distortion may not ensure full, solid-solution lithiation of a cubic phase, it is possible that this effect also delays the onset of the tetragonal distortion to higher Li content. Given the need to limit domain growth, and the role of Ti in doing so, reducing Ti-content may create larger domains and impede the ability of the $\delta$-phase to suppress the JT distortion and likely delay the two-phase reaction as well. We note here that other factors in the experimentally synthesized material such as the extent of disorder at the boundaries, spatial distribution of excess Li and Ti in the structure, the types of interfaces which result from the presence of Ti, the actual domain morphology, the domain size distribution, and the average domain size may also play a role in influencing the Spinel two-phase reaction. Future work should seek to assess the efficacy of engineering the domain structure to further improve the rate capability and energy content of the  $\delta$-phase, while preserving the stabilizing effect of the nanodomain structure.



\section{Conclusions}

In conclusion, we discuss through crystallographic theory and modelling that a transformation of disordered rock-salt to Spinel can result in 8 variants of Spinel. These distinct Spinel variants lead to a multi-domain structure in the $\delta$-phase. We show that boundaries between Spinel variants formed by exchanging the occupancy of $16c$ and $16d$ Spinel sub-lattices are generally highest in energy. We also identify that the lowest energy boundaries between different variants mostly form along the (100) plane. The majority of the boundaries reduce the low-migration barrier 0-TM channels that exist in fully-ordered Spinel, which is unfavorable for long-range Li-transport, but some of this effect is mitigated by the affinity of Li-excess to segregate to the domain boundaries. When Li-excess occupies the boundary and replaces Mn, additional 0-TM channels will be created. Hence, higher Li-content is expected to be favorable for enhancing rate capability in this multi-domain spinel structure. 

Ti also shows affinity to occupy the domain boundaries; however, due to its lower mobility as compared to Mn it is expected to influence the domain size by pinning the boundaries during domain growth. Due to high anti-site energy of Ti similar to that of Mn, the occupancy of the relatively sluggish Ti in the initial DRX structure will also influence the type of Spinel variant emerging around it. At the nanoscale domain sizes of the $\delta$-phase, the energy gain from preventing Ti in anti-sites may overcome the energy cost of forming higher energy boundaries, which in turn can result in an irregular domain morphology. Larger Ti-content may therefore result in smaller domains with more irregular boundaries forming between them; a structural feature which suppresses the tetragonal distortion in multi-domain lithiated Spinel nanostructure of the $\delta$-phase even if the domains are highly ordered and coherent. 

\begin{acknowledgement}

This work was supported by the Assistant Secretary for Energy Eficiency and Renewable Energy, Vehicle Technologies Ofice, under the Applied Battery Materials Program of the US Department of Energy (DOE) under contract number DE-AC02-05CH11231 (DRX+ program). This research used computational resources of the National Renewable Energy Laboratory (NREL) and the National Energy Research Scientific Computing Center (NERSC).


\end{acknowledgement}

\bibliography{acs}

\pagebreak

\section*{Supplementary information}

\renewcommand\thefigure{\thesection.\arabic{figure}}   
\setcounter{figure}{0}

\renewcommand\thetable{\thesection.\arabic{table}}
\setcounter{table}{0}

\makeatletter 
\renewcommand{\thefigure}{S\@arabic\c@figure}
\makeatother

\makeatletter
\renewcommand{\thetable}{S\@arabic\c@table}
\makeatother

\begin{figure}[hbt!]
\centering
\includegraphics[scale=0.25]{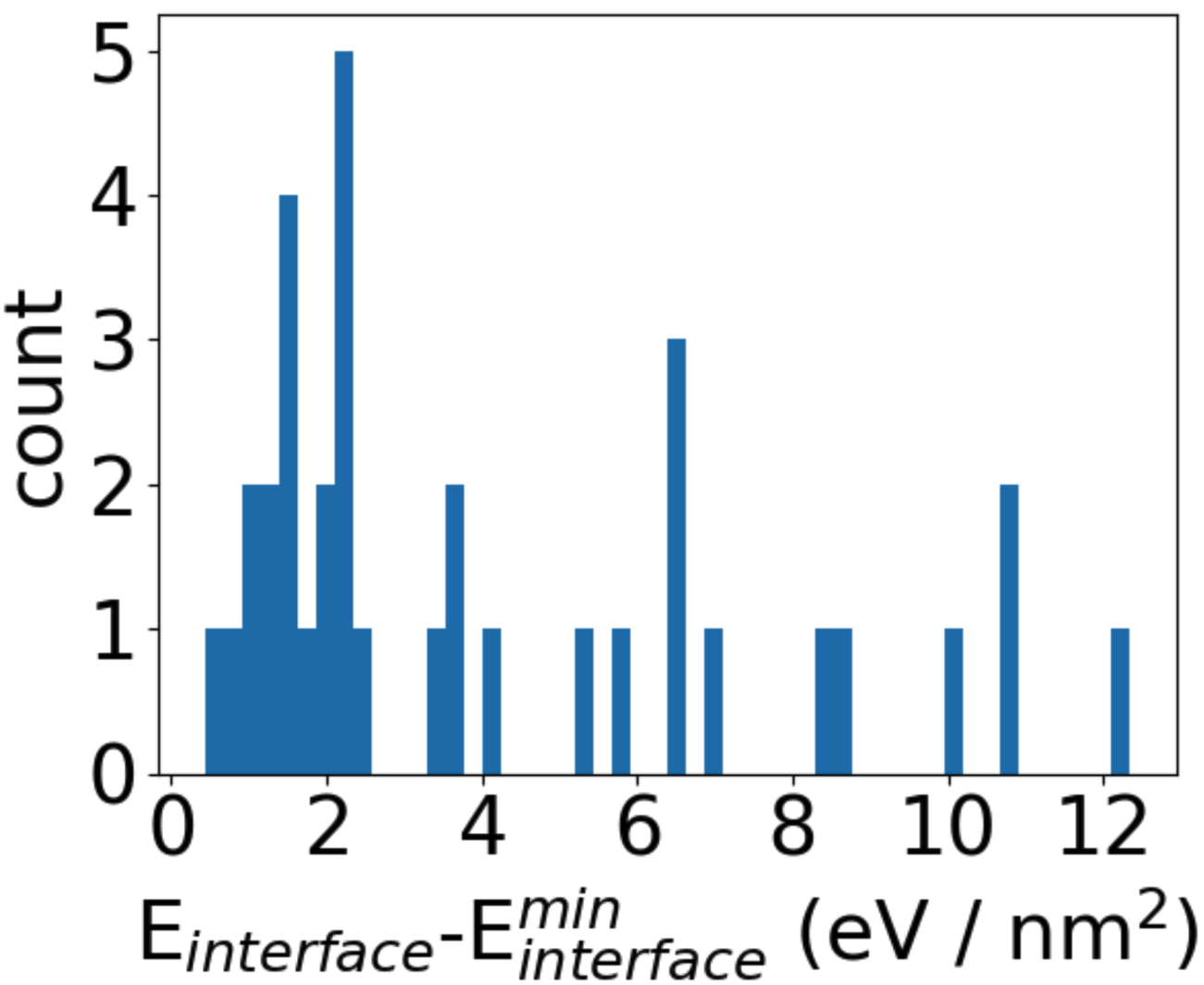}
\caption{Interfacial energy difference between the lowest energy interface and the 5 other higher energy interfaces formed by a combination of Spinel variants.}
\label{fig:Supp4}
\end{figure}

\end{document}